\documentclass[twocolumn,prl]{revtex4-1}
\usepackage{graphicx}    
\usepackage{dcolumn}     
\usepackage{bm}          

\begin{document}
\title{Model of the Electronic Structure of Electron-Doped Iron-Based Superconductors:
 Evidence for Enhanced Spin Fluctuations by Diagonal Electron Hopping}
\author{
Katsuhiro Suzuki$^1$, Hidetomo Usui$^1$, Soshi Iimura$^2$, Yoshiyasu Sato$^2$, \\
Satoru Matsuishi$^3$, Hideo Hosono$^{2,4}$, and Kazuhiko Kuroki$^1$}
\affiliation{$^1$ Department of Physics, 
Osaka University, 
Toyonaka, Osaka 560-0043, Japan}
\affiliation{$^2$ Materials and Structures Laboratory, 
Tokyo Institute of Technology, 
Yokohama 226-8503, Japan}
\affiliation{$^3$ Materials Research Center for Element Strategy, 
Tokyo Institute of Technology, 
Yokohama 226-8503, Japan}
\affiliation{$^4$ Frontier Research Center, 
Tokyo Institute of Technology, 
Yokohama 226-8503, Japan}
\date{\today}
\begin{abstract}
We present a theoretical understanding of the superconducting phase diagram of 
the electron-doped iron pnictides.
We show that, besides the Fermi surface nesting, 
a peculiar motion of electrons, where 
the next nearest neighbor (diagonal) 
hoppings between iron sites dominate over the nearest neighbor ones,  
plays an important role in the enhancement of the spin fluctuation 
and thus superconductivity.
In the highest $T_c$ materials, the crossover between the Fermi surface nesting 
and this ``prioritized diagonal motion'' regime occurs smoothly with doping, 
while in relatively low $T_c$ materials, the two regimes are separated and 
therefore results in a double dome $T_c$ phase diagram. 
\end{abstract}
\pacs{PACS numbers: }
\maketitle
In theoretical studies of high $T_c$ superconductors, one of the 
most important challenges is to extract the minimal essence of the material 
that leads to the strong pairing state. 
After the discovery 
of the iron-based superconductors\cite{Hosono}, 
the nesting between electron and hole Fermi surfaces has been 
considered as such an essential feature, 
and therefore, the identity of the family.  
In fact, several theoretical studies suggested 
a possibility of spin fluctuation mediated pairing, where the spin 
fluctuation arises around the nesting vector $(\pi,0)$
\cite{Mazin,Kuroki,Chubukov,Hirshfeld,Ikeda,Daghofer,Thomale,Wang}.
The spin fluctuation mediates $s\pm$-wave pairing, 
where the gap function has $s$-wave symmetry, 
but its sign is reversed between the electron and hole Fermi surfaces. 

However, recent experiments  
suggest that high $T_c$ is obtained when the nesting is degraded, 
or even in the absence of the 
nesting\cite{Iimura,KFe2Se2,KFe2Se2ARPES,STO,STO2}.
Then, a question of great interest is ``what is the key ingredient for high 
$T_c$  \textit{peculiar to the iron-based superconductors ?}'' In this context, 
the so-called hydrogen-doped 1111 systems, 
\textit{Ln}FeAsO$_{1-x}$H$_x$ (\textit{Ln} = Gd, Sm, Ce, La)\cite{Iimura} 
where a large amount of electrons can be doped by O~$\rightarrow$~H 
substitution,  provide us with some important clues. For \textit{Ln}~=~La, 
the $T_c$ vs $x$ (doping rate) phase diagram 
exhibits a double dome feature, and 
the second dome has higher $T_c$ than the first (see Fig.~\ref{fig2}). 
The normal state properties above $T_c$ 
such as the temperature dependence of the 
resistivity are also different between the two domes. 
On the other hand, for \textit{Ln}= Sm, Ce, Gd, the phase diagram 
exhibits a single dome feature, and very high $T_c$'s close to or 
exceeding 50 K are observed. These single dome materials share commonalities 
with the second $T_c$ dome of LaFeAsO$_{1-x}$H$_x$\cite{Iimura}, 
so that understanding the 
origin of the second dome directly leads to the origin of the 
very high $T_c$ in the iron-based superconductors.

One can easily expect that the Fermi surface nesting is degraded 
in the second dome compared to that in the first 
due to the large amount of doped electrons. In fact, the present study reveals 
that while the first $T_c$ dome originates from the spin fluctuation 
induced by the nesting of the Fermi surface having $d_{xz/yz}$ (and also 
$d_{xy}$ in some cases) orbital components, the second $T_c$ dome 
is due to the spin fluctuation enhanced by a peculiar motion 
of electrons within the $d_{xy}$, where the second neighbor diagonal 
hoppings are larger than the nearest neighbor ones. 
Such an electron motion is specific to the tetrahedral 
coordination of the pnictogen atoms, and 
we conclude that this prioritized diagonal motion is a 
key factor giving rise to the high $T_c$.
In the single dome $T_c$ materials, ``the nesting dominating'' 
and the ``prioritized diagonal motion'' regimes are not well separated, and the 
highest $T_c\sim 50$K is attained around the crossover regime. Then, 
another important key ingredient for high $T_c$ is that 
the $d_{xy}$ and $d_{xz/yz}$ orbitals both act as driving forces of the 
same pairing state, namely, $s\pm$-wave pairing.
Among various multiorbital systems, this is an 
unparalleled feature peculiar to the iron-based superconductors.

In 1111 systems,  electrons are doped into the FeAs layer 
by substituting O$^{2-}$ with F$^{-}$ or H$^{-}$.
The doping actually affects the electronic band structure in two ways; i.e., 
the increase of positive and negative charges in the 
\textit{Ln}O and FeAs layers, respectively, and the reduction of the 
As-Fe-As bond angle. The bond angle reduction occurs linearly with 
doping as shown in Fig.S1 of the Supplemental Material\cite{sup}. 
Changing the rare earth element appears as a 
parallel shift of the bond angle variance against the doping rate.
Quite recently, this trend has further been confirmed by 
partially replacing As by P in SmFeAsO$_{1-x}$H$_x$\cite{Matsuishi}, where 
a double dome phase diagram is found for sufficient amount of phosphorous 
substitution. 
To model these effects, 
band structure calculations are performed using the VASP code\cite{vasp} 
for hypothetical variations of 
LaFeAsO, where we (i) adopt 
the virtual crystal approximation replacing the oxygen potential 
by a $1-x:x$ mixture 
of oxygen and fluorine potentials\cite{Iimura} 
($x=0.05\sim 0.5$ with an increment of $0.05$), 
and (ii) vary the bond angle linearly 
according to $x$ as $\alpha(x)=-7.48x + 114.36 +\Delta\alpha$, 
where $\Delta\alpha$ is the 
amount of parallel shift made with respect to the 
the actual bond angle variance of LaFeAsO$_{1-x}$H$_x$.
We consider $\Delta\alpha=-3, -2, -1, 0, +1, +2^{\circ}$, 
as shown in Fig.S1 in the Supplemental Material\cite{sup}.
Varying $\Delta\alpha$  corresponds to considering 
materials with different rare earth (\textit{Ln})
or anion (As partially replaced by P) elements\cite{Lee}. 
To capture the essence, 
we vary only the bond angle, while fixing the Fe-As bond length.
By extracting the bands near the Fermi level using the 
Wannier90 package\cite{Wannier90}, 
we construct 
models consisting of 
$d_{xy}$, $d_{yz}$, $d_{xz}$, $d_{x^2-y^2}$, and $d_{3z^2-r^2}$ 
Wannier orbitals\cite{Kuroki}. Considering that the three dimensionality 
is not essential to the single vs. double dome issue, we 
omit the interlayer hoppings, and concentrate on two dimensional models  
in which the Brillouin zone can be unfolded to obtain a five orbital model
\cite{Kuroki}.

\begin{figure}[t]
\centering
\includegraphics[width=9cm,clip]{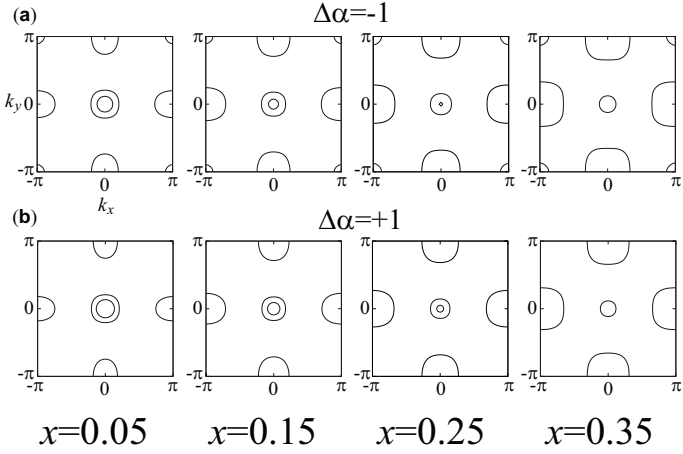}
\caption{
Fermi surfaces for $x=0.05$, $x=0.15$, $x=0.25$, $x=0.35$
with (a) $\Delta\alpha=-1^{\circ}$ or (b) $\Delta\alpha=+1^{\circ}$.
\label{fig1}}
\end{figure}

\begin{figure}[t]
\centering
\includegraphics[width=9cm,clip]{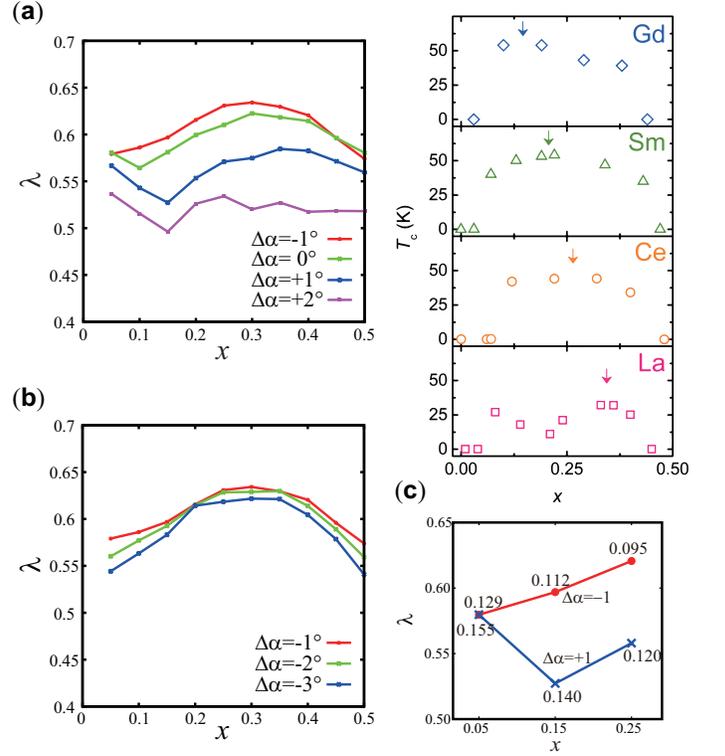}
\caption{
Eigenvalue of the Eliashberg equation against doping.
(a) $\lambda$ against doping for $\Delta\alpha=-1$, $0$, $+1$, $+2^{\circ}$. 
(b) Similar as in (a) for $\Delta\alpha=-3$, $-2$, $-1^{\circ}$.
(c) $\lambda$ 
for simplified models in which only $t_1$ is varied in conjunction with 
$x$ so as to maintain the 
volume of the $(\pi,\pi)$ Fermi surface. The numbers are the 
values of $t_1$ (in eV). Here, $t_2$ is fixed at 0.106 (0.113) eV 
for $\Delta\alpha=-1$ $(+1)$. See text for more details.
Upper right panel : 
Experimental result of $T_c$ vs $x$  for \textit{Ln}FeAs(O,H) with 
\textit{Ln}=Gd, Sm, Ce, and La (from Ref.~\onlinecite{Iimura}).}
\label{fig2}
\end{figure}

Figure~\ref{fig1} shows the Fermi surface evolution
with doping  for $\Delta\alpha=+1^{\circ}$ and $-1^{\circ}$. 
The main difference between the two cases is the presence or absence of the 
Fermi surface around the wave vector $(\pi, \pi)$, which originates 
from the $d_{xy}$ orbital\cite{KurokiPRB}.
The volume of the electron Fermi surfaces around $(\pi, 0)$ and 
$(0, \pi)$ increases with doping, and the hole Fermi surfaces around 
$(0, 0)$, arising from the $d_{xz}/d_{yz}$ orbitals, shrink.
On the other hand, The volume of the $d_{xy}$ hole Fermi surface around $(\pi, \pi)$ 
remains nearly unchanged due to the band structure variation with doping\cite{SuzukiJPSJ,Yamakawa}. 
In any case, the volume difference between electron and hole Fermi surfaces 
increases with doping, so that the nesting  
becomes ill-conditioned.

Considering intra- and interorbital
electron-electron interactions 
on top of the five orbital band structure, 
we apply the fluctuation exchange (FLEX) 
approximation\cite{Ikeda,Bickers,Dahm} to each model, 
and obtain the eigenvalue of the Eliashberg equation $\lambda$ (at a fixed 
temperature of $T=0.005$eV), which is taken as a measure of $T_c$. 
We take intraorbital $U=1.3$eV, the interorbital $U'=U-2J$, 
Hund's coupling, and the pair hoppings $J=J'=U/6$.
In a previous study, we adopted random phase approximation, 
where the self energy correction was neglected\cite{SuzukiJPSJ}. 
There, the eigenvalue of the Eliashberg equation was found to be 
monotonically enhanced 
with electron doping, which does not agree with the experimental observations. 
Also, the origin of the material dependence of the phase diagram was 
not clarified.

The calculated eigenvalues of the Eliashberg equation $\lambda$ 
for $\Delta\alpha=-1^{\circ}\sim +2^{\circ}$ are shown in Fig.~\ref{fig2}(a).
For $\Delta\alpha=-1^{\circ}$, the $\lambda$ against 
$x$ plot shows a ``single dome'' variance. This is already quite 
interesting in that the magnitude of $\lambda$ (and, hence, $T_c$) 
is maintained in such a large doping range.
Even more interestingly, for $\Delta\alpha=0^{\circ}$, there 
appears a slight dip in the lightly doped regime, and this feature 
becomes more pronounced for $\Delta\alpha=+1^{\circ}$ and $+2^{\circ}$. 
Also, the maximum $T_c$ is obtained at a larger doping rate 
when $\Delta\alpha$ is increased.
Similar results are obtained also for orbital dependent interactions 
(Supplemental Material\cite{sup} Fig.~S3).
In Fig.~\ref{fig2}~(b), we show the doping dependence of $\lambda$ 
for $\alpha=-1^{\circ}\sim -3^{\circ}$. 
It can be seen that the increase of $\lambda$ with $x$ in the lightly doped 
regime becomes more rapid with decreasing $\Delta \alpha$. 
These results are overall 
in good agreement with the trend observed experimentally 
in Ref.~\onlinecite{Iimura} (Fig.~\ref{fig2}, upper right panel) and also 
Ref.~\onlinecite{Lee}.

\begin{figure}[t]
\centering
\includegraphics[width=8cm,clip]{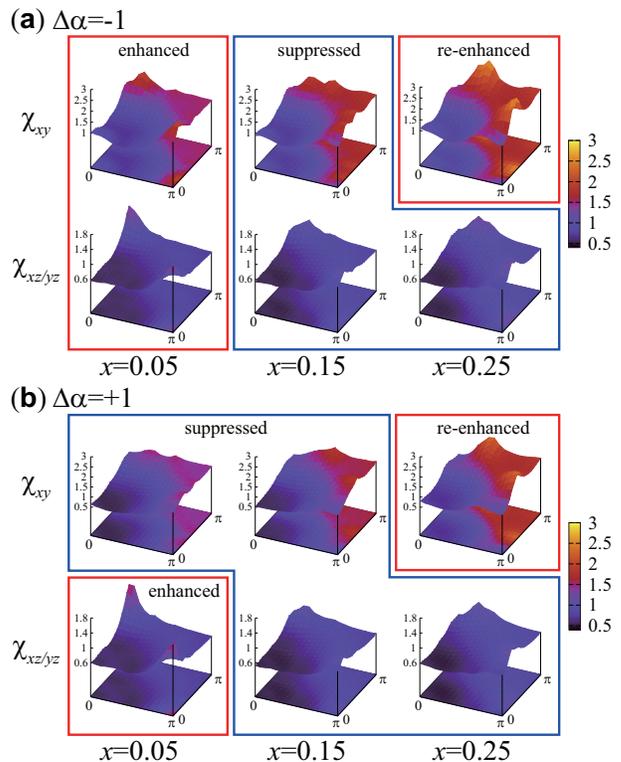}
\caption{
Intraorbital spin susceptibilities $\chi_{xy}$ and $\chi_{xz/yz}$ 
for $x=0.05$, 0.15, and 0.25.
(a) $\Delta\alpha=-1^{\circ}$ and (b) $\Delta\alpha=+1^{\circ}$.}
\label{fig3}
\end{figure}

In Fig.~\ref{fig3}, we show the doping dependence of the 
intraorbital spin susceptibility $\chi_{xy}$ and $\chi_{xz/yz}$ 
within the $d_{xy}$ and $d_{xz}/d_{yz}$ orbitals\cite{Comment}  
for $\Delta\alpha=+1^{\circ}$ and $-1^{\circ}$. 
Let us first focus on $\Delta\alpha=-1$. 
For $x=0.05$, there appear peaks around $(\pi,0)$ and 
$(0,\pi)$ in both $\chi_{xy}$ and $\chi_{xz/yz}$ reflecting the 
Fermi surface nesting in the lightly doped system. 

These peak structures are suppressed 
by electron doping because the nesting is degraded.
However, $\chi_{xy}$ is unexpectedly reenhanced beyond $x\sim 0.2$.
The reason for this cannot be the Fermi surface nesting in its original 
sense because the 
nesting is monotonically degraded by doping.
For $\Delta\alpha=+1^{\circ}$, on the other hand, the variance of $\chi_{xy}$ is 
different in that there is no enhancement in the lightly doped regime. 
The absence of enhanced $\chi_{xy}$ there 
is natural because the $d_{xy}$ hole Fermi surface 
around $(\pi,\pi)$ is absent for $\Delta\alpha=+1^{\circ}$, 
so that  there is no Fermi surface nesting. Conversely, 
it is surprising to find an enhancement in the largely doped regime.
Interestingly, an inelastic neutron scattering 
experiment for LaFeAsO$_{1-x}$H$_x$ actually observes 
the suppression of the spin fluctuation around $x\sim 0.2$ and 
its reenhancement in the largely doped regime\cite{IimuraPRB}.
Also, comparing $\Delta \alpha=-1^{\circ}$ and $\Delta \alpha=+1^{\circ}$, 
the spin fluctuation grows more 
rapidly with doping for the former than for the latter.

\begin{figure}[t]
\centering
\includegraphics[width=7cm,clip]{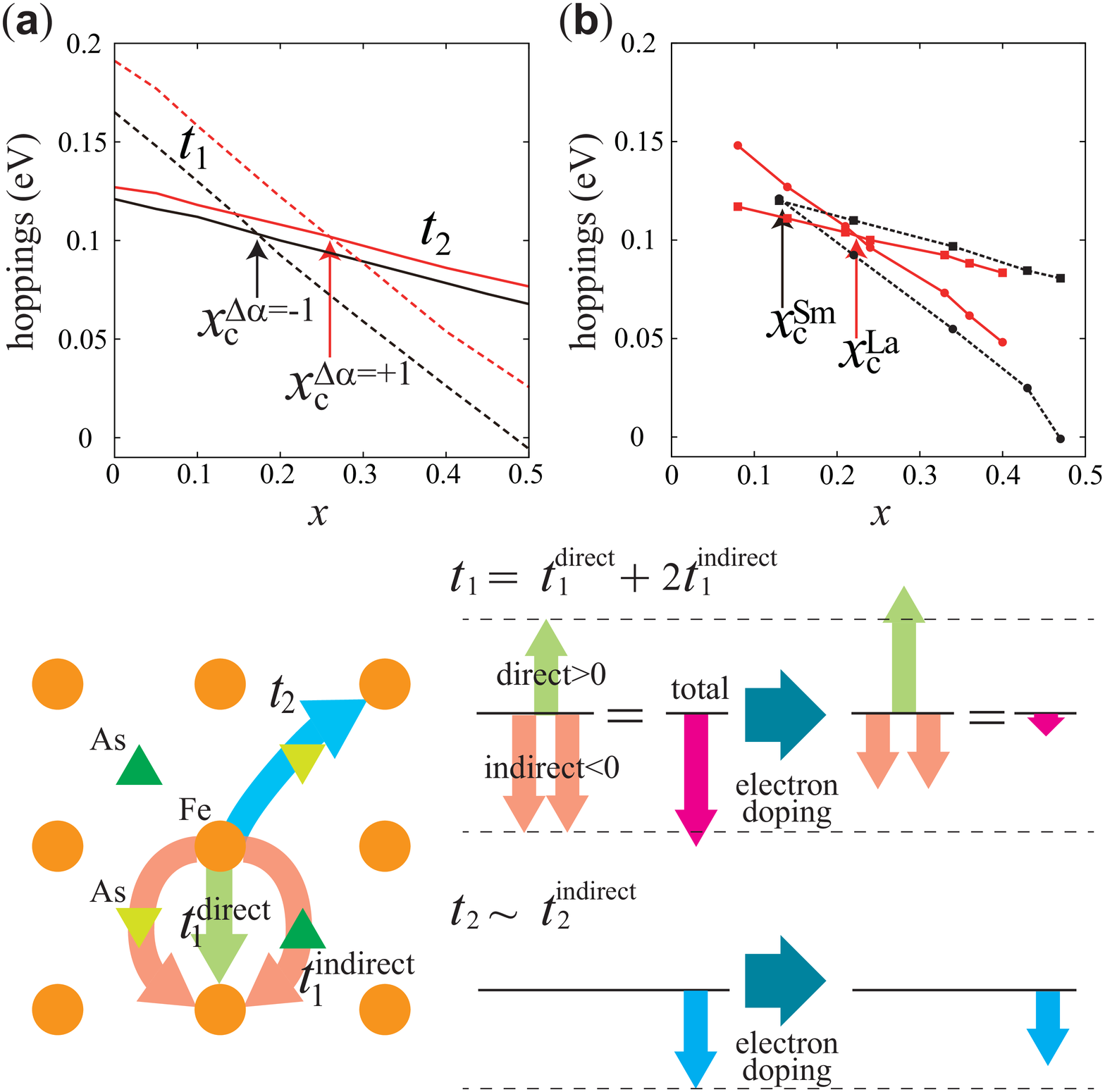}
\caption{
Upper panels: doping dependence of the real space hopping integrals
$t_1$ and $t_2$ against the doping. 
(a) $\Delta\alpha=+1^{\circ}$ (red or gray) and $\Delta\alpha=-1^{\circ}$ (black).
Solid (dashed) lines are $t_2$ ($t_1$). 
(b) Actual materials La1111(red or gray) and Sm1111(black). Circles (boxes) 
are $t_1$ ($t_2$). 
Lower panels : 
schematic figure of the origin of the prioritized diagonal hopping 
with doping.
\label{fig4}}
\end{figure}

The doping dependence of the Eliashberg equation 
eigenvalue $\lambda$ and the intraorbital spin fluctuations 
are strongly correlated, 
so that understanding the latter directly leads to 
the understanding of the former.
Since the Fermi surface evolution does not seem to be correlated with the 
doping dependence of the spin fluctuation, 
we now focus on the real space hopping integrals within 
the $d_{xy}$ orbitals. In 
Fig.~\ref{fig4}, we plot the doping dependence of the nearest $(t_1)$ and the 
next nearest 
$(t_2)$ neighbor hoppings 
within the $d_{xy}$ orbitals for $\Delta\alpha=-1^{\circ}$ 
and $+1^{\circ}$. The nearest neighbor hopping $t_1$ decreases rapidly 
with doping, and becomes smaller than $t_2$ at a certain 
doping rate $x_c\sim 0.17$ for 
$\Delta\alpha=-1^{\circ}$ and $x_c\sim 0.28$ for $\Delta\alpha=+1^{\circ}$. 
We also show in Fig.~\ref{fig4}(b) the calculation result for the 
actual La1111 and Sm1111 
using the experimentally determined lattice parameters.
It is indeed seen that $x_c$ is larger for La than for Sm corresponding 
to the larger bond angle in the former.
We will refer to this peculiar hopping relation $t_2>t_1$ as  
``prioritized'' diagonal motion (or hopping) of electrons.

The rapid decrease of $t_1$ by doping as compared to $t_2$ 
can be understood as a combined 
effect of  (i) the increased positive charge in the LaO layer, 
(ii) reduction of the Fe-Fe distance, and (iii) increase of the pnictogen 
height, where (ii) and (iii) are the effects of the bond angle reduction.
In the five orbital model, we consider Wannier orbitals, which 
implicitly take into account the Fe $3d$ and the hybridized 
As $4p$ atomic orbitals.
If we consider these atomic orbitals explicitly, the present 
$t_1$ can be expressed as 
$t_1=t_1^{\rm direct}+2t_1^{\rm indirect}$, where 
$t_1^{\rm direct}$ and $t_1^{\rm indirect}$ are contributions from the direct hopping  
between Fe 3$d_{xy}$ orbitals and the indirect hopping via As $4p$, 
respectively, as shown in Fig.~\ref{fig4}. 
The two contributions have opposite signs, and $t_1^{\rm indirect}$ dominates 
in the lightly doped regime. This cancellation of the 
direct and indirect hoppings has been discussed in Refs.\cite{Miyake42622,Kotliar}. On the other hand, the next nearest neighbor $t_2$ is mainly governed 
by $t_2^{\rm indirect}$  because of the larger Fe-Fe distance.
As electrons are doped, the energy level of the As $4p$ orbital is lowered and 
moves away from the Fe $3d$ level due to the effect of (i), so that the indirect 
hoppings decrease. The indirect contribution is also reduced because of (iii). 
By contrast, the direct hopping increases due to (ii). 
The combined effect of increasing $t_1^{\rm direct}$ and 
decreasing $|t_1^{\rm indirect}|$ results in a rapid decrease of $t_1$  with doping. 
The effect is weak for $t_2$ because it is mainly dominated by $t_2^{\rm indirect}$.
As can be understood from this explanation, $x_c$ is larger for materials with larger $\Delta\alpha$. 

Intuitively, $t_1<(>)t_2$ corresponds to $J_1<(>)J_2$ in the limit of strong 
electron correlation\cite{Si,Hu} since $J_i\propto t_i^2/U$, where 
$U$ is the on-site intraorbital repulsion. 
Therefore, $t_1 < [ > 0]t_2$ is naively 
expected to be in favor of the $(\pi, 0)$ [$(\pi, \pi)$]
spin fluctuations. More precisely, however, 
the enhancement of the spin fluctuation 
in the largely doped regime should be traced back to the band structure 
(not just the Fermi surface)  since we are 
adopting FLEX, 
which is essentially a weak coupling approach.
In fact, as shown in the Supplemental Material\cite{sup} (Fig.S2), 
the shape of the band 
changes with doping, which is mainly due to the reduction of $t_1$. 
The disappearance of the van Hove singularity around the 
wave vector $(\pi,0)$ (reminiscent of those commonly seen in the cuprates) 
works in favor of the $(\pi,0)/(0,\pi)$ spin fluctuations over 
$(\pi,\pi)$.

In the models adopted above, not only $t_1$, but also other parameters 
vary with electron doping. To more directly single out the reduction of 
$t_1$ as the key factor, we have done the following analysis using 
simplified models.
Namely, we start with five orbital models derived from a first principles 
band calculation performed with 
 $\Delta\alpha = -1$ or +1, both with $15\%$ fluorine doping.
Within these models, we vary the electron density 
in the range of $0.05 \leq x \leq 0.25$. If all the hoppings were 
fixed (rigid band), the hole Fermi surface would monotonically shrink
as $x$ increases. As seen above, however, 
the Fermi surface around $(\pi,\pi)$ 
is almost unchanged with electron doping if the  variance of the 
lattice parameters and the O~$\rightarrow$~F substitution 
effect is taken into account in the first principles calculation. 
To simulate this effect, in the simplified models, 
we vary only $t_1$ by hand in conjunction with the 
electron doping so that the volume of the 
$(\pi,\pi)$ Fermi surface 
is the same as that for the original model. 
For $\Delta\alpha=+1$, where the $(\pi,\pi)$ hole Fermi surface is absent,  
the energy difference between the chemical potential and the 
top of the $d_{xy}$ band at $(\pi,\pi)$ is kept to be the same as 
that in the original model. $\lambda$ calculated this way as a function  of 
$x$ for $\Delta\alpha=\pm 1$ is shown in Fig.~\ref{fig2}~(c), where 
a trend similar to that in Fig.~\ref{fig2}~(a) is seen; 
when $t_1$ is significantly larger than $t_2$, 
$\lambda$ decreases with doping, while  when $t_1$ is comparable to 
or smaller than $t_2$, 
$\lambda$ increases with doping. 

\begin{figure}[t]
\centering
\includegraphics[width=7cm,clip]{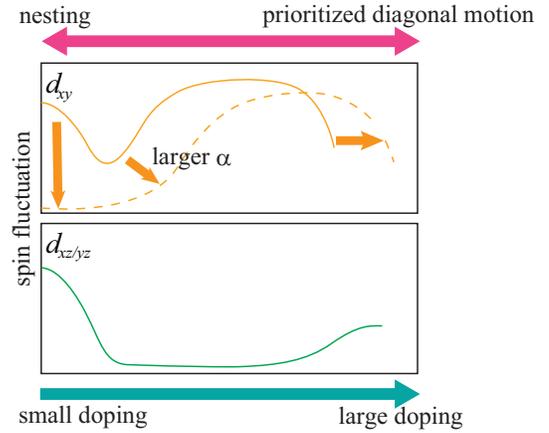}
\caption{Schematic figure of the spin fluctuation contribution to superconductivity.
\label{fig5}}
\end{figure}

In Fig.~\ref{fig5}, we show a schematic figure of the spin fluctuation 
contribution to $s\pm$ superconductivity.
For the $d_{xz}/d_{yz}$ orbital, there is 
spin fluctuation mediated pairing arising from 
good nesting in the lightly doped regime, which is suppressed by doping 
because the nesting is degraded.
In the $d_{xy}$ orbital, 
there can be moderate Fermi surface nesting  
in the lightly doped regime depending on the absence or presence of the 
$d_{xy}$ hole Fermi surface around $(\pi, \pi)$.
Therefore, for materials with small (i.e., negative) $\Delta\alpha$, 
the $d_{xy}$ spin fluctuation  crosses over from the 
nesting to the prioritized diagonal motion regime. 
On the other hand, in materials with large (positive) $\Delta \alpha$, 
there is no nesting regime in the $d_{xy}$ orbital, so that the $d_{xy}$ spin 
fluctuation monotonically increases with doping.
For materials with small 
bond angle, the crossover from the nesting 
to the prioritized diagonal motion regime 
occurs smoothly because $x_c$ is small. Therefore, the $T_c$ phase diagram 
consists of a single dome. $x_c$ is large 
for materials with large bond angle, so that the two regimes are 
separated, resulting in a 
double dome structure of the phase diagram. 
Interestingly, we have also come to realize a relation between the 
spin fluctuation and the resistivity, 
which is explained in the Supplemental Material\cite{sup} (Fig.S4).

To conclude, our study has revealed the importance of the 
peculiar motion of electrons in the $d_{xy}$ orbital, especially in 
cases with very high $T_c$. 
Further tests for the present conclusion can be performed by 
examining the pressure effect. In Ref.\cite{Iimura}, 
it was found that applying pressure to LaFeAs(O,H) makes 
the double dome $T_c$ phase diagram turn into a single dome one. 
Our preliminary 
theoretical study on this problem shows 
that applying pressure enhances the $t_2/t_1$ 
ratio, and hence has an effect similar to that of replacing La by, say, Ce. 
Detailed analysis on this problem will be presented elsewhere. 
Also, it would be interesting to experimentally investigate 
\textit{Ln}FeAs$_{1-y}$P$_y$O$_{1-x}$H$_x$
other than \textit{Ln}=Sm\cite{Matsuishi} as another test for the 
present conclusion.
A surprisingly interesting feature of the iron-based superconductors 
is that the prioritized diagonal motion in the $d_{xy}$ orbitals and 
the nesting within $d_{xy}$ or $d_{xz/yz}$ Fermi surfaces can all 
be driving forces of the $s\pm $-wave superconductivity. 
This coherent cooperation among various components is indeed the 
unparalleled identity of the iron-based superconductors.

We thank H. Sakakibara, S. Onari, Y. Yamakawa, and R. Arita for valuable discussions.
Numerical calculations were performed at the facilities of the Supercomputer Center,
Institute for Solid State Physics, University of Tokyo.
This study has been supported by Grants-in-Aid  for Scientific Research 
No.~24340079 and No.~25009605 from the Japan Society for the Promotion of Science. 
The part of the research at the Tokyo Institute of Technology was supported by the JSPS FIRST Program.



\begin{thebibliography}{99}
\bibitem{Comment} The actual calculation is done with the orbital basis 
that are rotated by 45$^{\circ}$ from the Fe-Fe direction $x$ and $y$, 
namely, $X$ and $Y$ defined in Ref.~\onlinecite{Kuroki}.
Here, the $YZ$ component of the intraorbital spin susceptibility is adopted.
\end{thebibliography}
\end{document}


\title{Model of the Electronic Structure of Electron-Doped Iron-Based Superconductors:
 Evidence for Enhanced Spin Fluctuations by Diagonal Electron Hopping}
\author{
Katsuhiro Suzuki$^1$, Hidetomo Usui$^1$, Soshi Iimura$^3$, Yoshiyasu Sato$^3$, \\
Satoru Matsuishi$^4$, Hideo Hosono$^{3,5}$, and Kazuhiko Kuroki$^1$}
\affiliation{$^1$ Department of Physics, 
Osaka University, 
Toyonaka, Osaka 560-0043, Japan}
\affiliation{$^3$ Materials and Structures Laboratory, 
Tokyo Institute of Technology, 
Yokohama 226-8503, Japan}
\affiliation{$^4$ Materials Research Center for Element Strategy, 
Tokyo Institute of Technology, 
Yokohama 226-8503, Japan}
\affiliation{$^5$ Frontier Research Center, 
Tokyo Institute of Technology, 
Yokohama 226-8503, Japan}
\date{\today}

\maketitle
\noindent
\section{Method}
\begin{figure}
\centering
\includegraphics[width=8cm,clip]{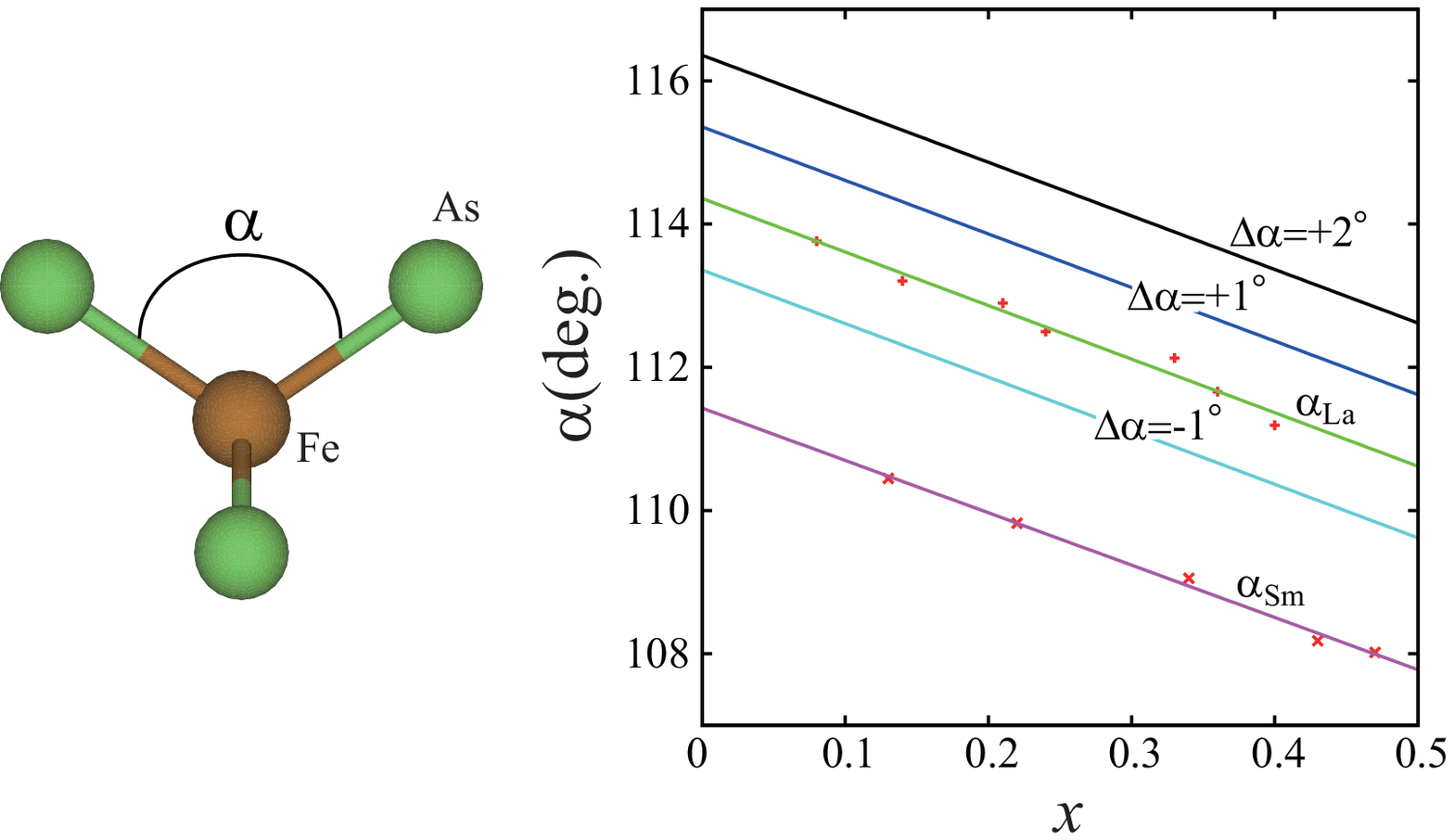}
\caption{Doping dependence of the As-Fe-As  bond angle. The dots are the 
actual experimental data.}
\label{figS1}
\end{figure}

Here we describe the details of the adopted method. 
In the actual materials, the bond angle reduction occurs 
almost linearly with doping as shown as $\alpha_{\rm La}$ 
and $\alpha_{\rm Sm}$ in Fig.~\ref{figS1}.
In our calculation, we vary the bond angle linearly 
according to $x$ as $\alpha(x)=-7.48x + 114.36 +\Delta\alpha$, 
where $\Delta\alpha$ is the amount of parallel shift made 
with respect to the bond angle variance of the actual La1111.

First principles band calculation is performed by 
using the VASP package\cite{vasp}. 
We consider hypothetical materials LaFeAsO$_{1-x}$F$_x$,  
where the As-Fe-As bond angle is determined as mentioned in the main text.
Values of all the other lattice parameters are those determined experimentally 
for LaFeAsO$_{0.92}$H$_{0.08}$. The effect of partially substituting O by F 
is taken into account through the virtual crystal approximation.
Here, we adopt GGA-PBEsol exchange-correlation functional\cite{PBEsol}. 
The wave functions are expanded by plane waves up to 
cut-off energy of 550eV and 10$^3$ $k$-point meshes are used.
Five orbital tight-binding models are derived from the 
first principles band calculation exploiting the maximally localized 
Wannier orbitals\cite{Vanderbilt}. The Wannier90 code is used for 
generating the Wannier orbitals\cite{Wannier90}.
We neglect the hopping in the $z$ direction and concentrate on the 
two dimensional model for simplicity.

We consider two sets of electron-electron interactions, 
orbital independent and dependent ones.
In standard notations, the orbital independent interactions are taken 
as follows : the intraorbital interaction is $U=1.3$eV, the interorbital $U'=U-2J$, 
Hund's coupling and the pair hoppings $J=J'=U/6$.
The orbital dependent interactions are obtained by multiplying a 
constant factor (0.55 here) to all of the interactions 
obtained for LaFeAsO in Ref.~\onlinecite{Miyake}. 
The results for the orbital independent interactions are shown in the 
main text, while those for the orbital dependent interactions are shown 
below.

We apply the FLEX\cite{Bickers} 
method to the obtained five orbital model. In FLEX, 
bubble and ladder type diagrams are collected to calculate the spin 
$\chi^S_{ijkl}(\Vec{q},i\omega_n)$ and 
charge susceptibilities $\chi^C_{ijkl}(\Vec{q},i\omega_n)$, 
from which the self energy is calculated. 
Here, $i,j,k,l$ are orbital indices, and the 
intraorbital susceptibilities displayed in the main text are obtained as 
$\chi_{iiii}(\Vec{q},i\omega_n=0)$. 
The renormalized Green function is obtained from the Dyson equation.
This process is repeated until a self consistent solution is obtained.
As was done in Ref.~\onlinecite{Ikeda}, 
we subtract the $\omega=0$ component of the 
self energy for each iterations, 
which is considered to be taken into account already in the 
first principles band calculation. 
The resulting Green function and the susceptibilities are plugged into the 
linearized Eliashberg equation, whose eigenvalue $\lambda$ reaches unity 
at $T=T_c$, the superconducting transition temperature. 
Instead of going down to $T_c$, which is generally a tedious calculation, 
we obtain $\lambda$ at a fixed temperature ($T=0.005$eV here), and use it 
as a qualitative measure for $T_c$. 
In the FLEX calculation, 
we take $32\times 32$ $k$-point meshes and 4096 Matsubara frequencies. 
We also perform calculations for $64\times 64$ 
$k$-point meshes, or 8192 Matsubara frequencies for comparison 
for some of the parameter sets.
\begin{figure*}
\centering
\includegraphics[width=12cm,clip]{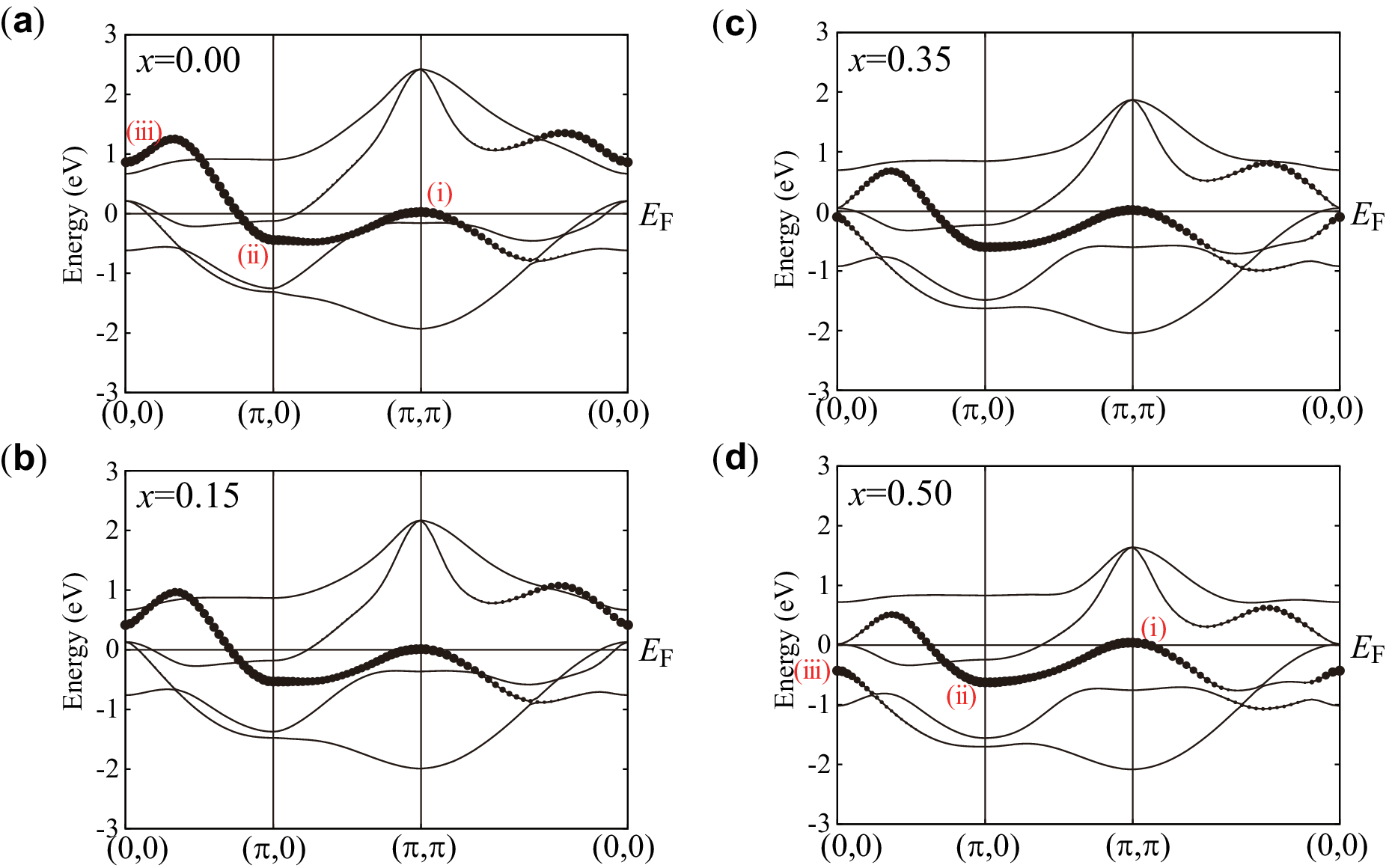}
\caption{Band structure variance against doping for $\Delta \alpha=0^{\circ}$.
(a) $x=0.0$, (b) $x=0.15$, (c) $x=0.35$, (d) $x=0.5$. 
The thickness represents the strength of the $d_{xy}$ orbital character.
The $d_{xy}$ portion of the band around $(\pi,\pi)$, $(\pi,0)$, and $(0,0)$ 
are denoted as (i), (ii), and (iii), respectively.}
\label{figS2}
\end{figure*}

\section{Band structure variation with doping} 
The band structure varies with doping as shown in Fig.~\ref{figS2}, 
which is mainly due to the reduction of $t_1$.
One effect is that the $d_{xy}$ portion 
around $(\pi,\pi)$ is pushed up, which cancels 
out with the increase of the electron number to result in an unchanged 
Fermi surface.
Another variation with doping occurs around the wave vector $(\pi,0)$. 
For small doping rate (large $t_1$), a saddle point is present in the 
$d_{xy}$ band around the 
wave vector $(\pi, 0)/(0, \pi)$, which gives a diverging  density of states.
Although this van Hove singularity itself is 
somewhat away from the Fermi level, it does affect the 
spin susceptibility, which is obtained by summing up contributions 
including those away from the Fermi level.
This kind of band shape is reminiscent of the high $T_c$ cuprates, 
where the spin fluctuation is enhanced around $(\pi, \pi)$.  
On the other hand,  the saddle point disappears (actually it moves away 
from $(\pi, 0)$) for small $t_1$, 
and becomes a local minimum. Such a band shape no longer enhances the 
$(\pi,\pi)$ spin fluctuation, and therefore has better matching with the 
Fermi surface configuration 
with holes and electrons around $(\pi, \pi)$ and $(\pi, 0)/(0, \pi)$, 
respectively, which itself is in favor of the $(\pi, 0)$ spin fluctuations.
Since $x_c$ for $\Delta\alpha=-1$ is smaller than for 
$\Delta\alpha=+1$, the difference between $\sim(\pi,0)$ and $(\pi,\pi)$ spin 
fluctuations grows faster with doping for $\Delta\alpha=-1$, as seen 
in Fig.~4 of main text. 
The present analysis shows the importance of the prioritized diagonal motion of 
the electrons for the enhanced $\sim (\pi,0)$ spin 
fluctuation and thus $s\pm$ superconductivity.  

\section{Additional analysis regarding the eigenvalue of the Eliashberg 
equation  and the spin fluctuation}
\begin{figure}
\centering
\includegraphics[width=5cm,clip]{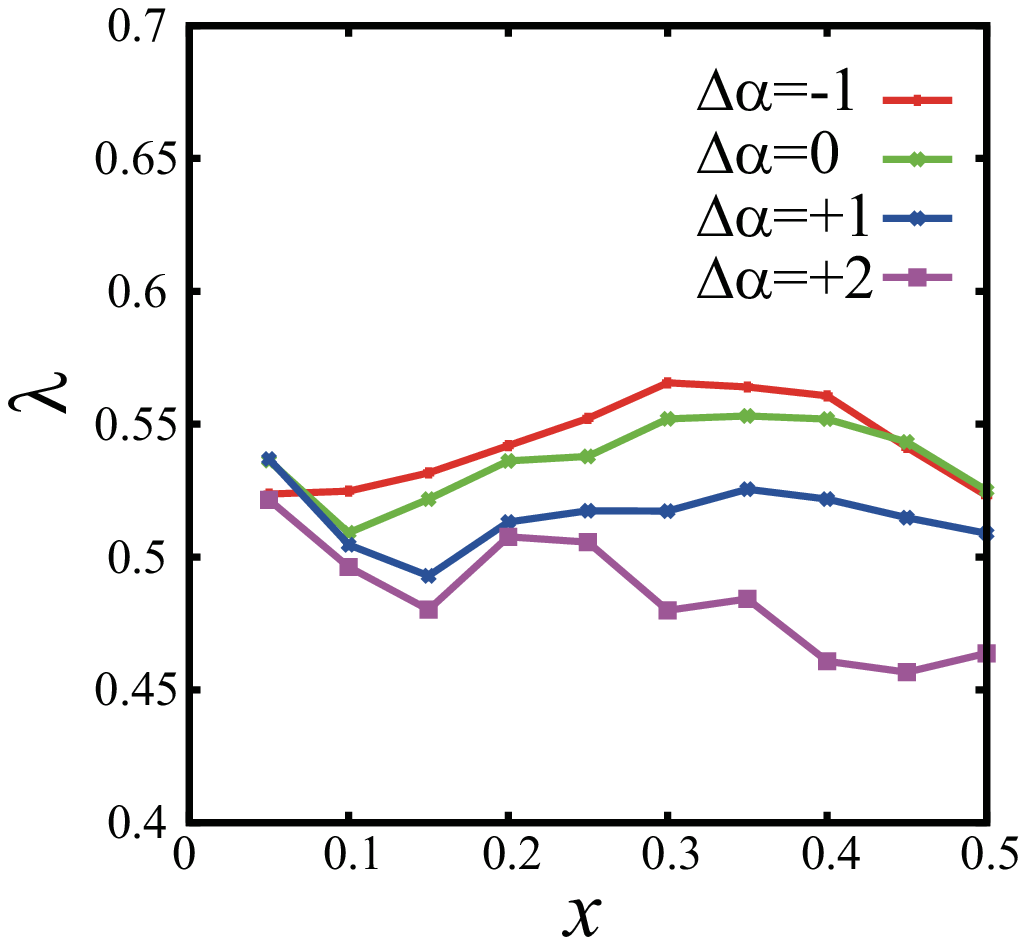}
\caption{Eigenvalue of the Eliashberg equation against doping 
for orbital dependent electron-electron interactions.}
\label{figS3}
\end{figure}
We show in Fig.~\ref{figS3} the doping dependence of $\lambda$ 
obtained by adopting orbital dependent electron-electron interactions  
proportional to those evaluated in Ref.~\onlinecite{Miyake} for LaFeAsO. 
It can be seen that the overall trend is the same 
as in the case for the orbital independent interactions presented in the 
main text.

In the main text, we have not put much focus 
on the variance of $\lambda$ against 
$\Delta \alpha$ (corresponding to the rare earth element dependence of $T_c$) 
for a fixed doping rate. As seen in Fig.2 of the main text (and also 
partially Fig.~\ref{figS3} here), 
$\lambda$ first increases with decreasing $\Delta\alpha$, 
then tends to saturate, and finally is reduced for smaller $\Delta\alpha$.
This is similar to the trend observed experimentally\cite{Lee} as well 
as the results obtained in a previous theoretical study\cite{Usuiangle}. 
In Ref.~\onlinecite{Usuiangle}, the reduction of the number of hole 
Fermi surfaces in the small $\Delta\alpha$ regime
has been proposed as the origin of the $T_c$ suppression there, 
but the more detailed present analysis 
shows that the $\lambda$ suppression occurs even when the 
Fermi surface multiplicity is maximized.
Within the present study,
the maximum $\lambda$ is attained when the $(\pi, \pi)$ 
$d_{xy}$ band is touching the Fermi level, i.e., when the $d_{xy}$ Fermi surface 
nesting appears to be ill-conditioned. In addition to $t_2 > t_1$, 
this condition also has to be fulfilled for the highest $T_c$. 
This is probably because the 
finite energy spin fluctuation that is responsible for pairing 
is enhanced when the top of the $d_{xy}$ hole band is touching the 
Fermi level. More 
detailed analysis on this issue is underway.

\begin{figure}
\centering
\vspace{1.3em}
\includegraphics[width=9cm,clip]{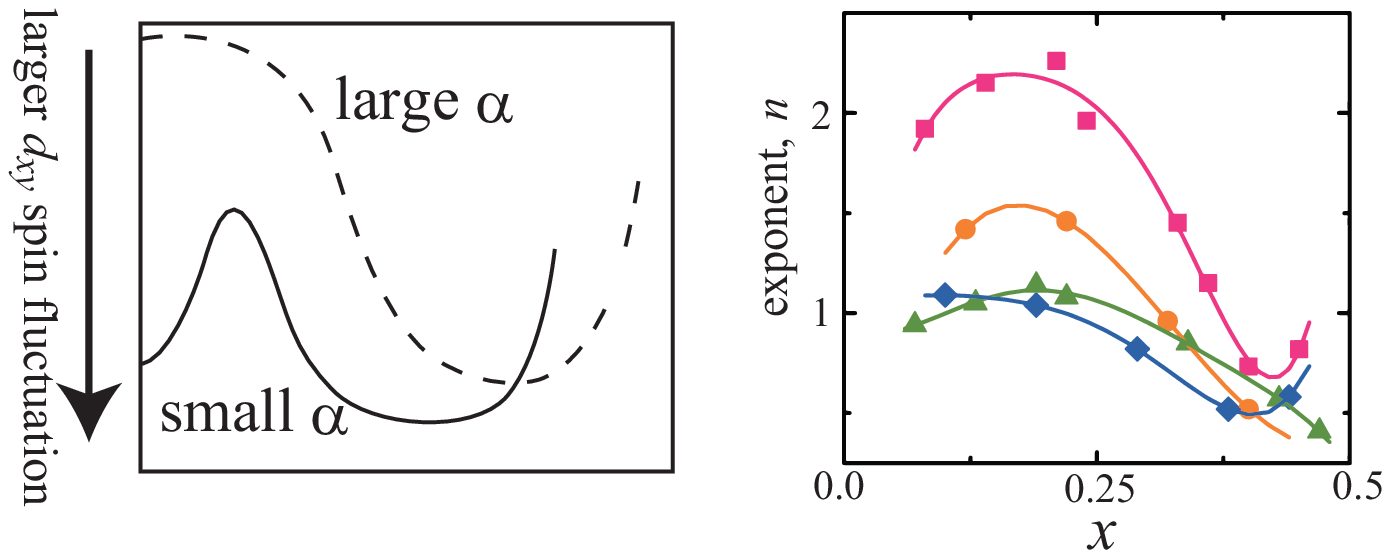}
\caption{Left panel : a figure obtained by turning upside down the 
upper panel of Fig.5 in the main text.
Right panel : the experimentally observed doping dependence of the 
$n$, where $n$ is the exponent of the temperature dependence of the 
resistivity $\rho\sim T^n$ (taken from Ref.~\onlinecite{Iimura}).}
\label{figS4}
\end{figure}

Regarding the doping dependence of the spin fluctuation, 
it is interesting to note that if we turn up side down the $d_{xy}$ orbital 
curves in Fig.~\ref{figS4}, they are reminiscent of the $n$ vs. $x$ relation 
observed experimentally (right panel of Fig.~\ref{figS4})\cite{Iimura}, 
where $n$ is the exponent of the temperature 
dependence of the resistivity. This means that the exponent is probing the 
strength of the spin fluctuation within the $d_{xy}$ orbital. This is natural 
since the portion of the Fermi surface with light mass determines the 
transport properties, and the light portion (the portion with large 
group velocity) is indeed 
the $d_{xy}$ orbital part  of the electron Fermi surface.

\section{Re-enhancement of the $\bm{d_{xz}/d_{yz}}$ spin fluctuation}
\begin{figure}
\centering
\includegraphics[width=9cm,clip]{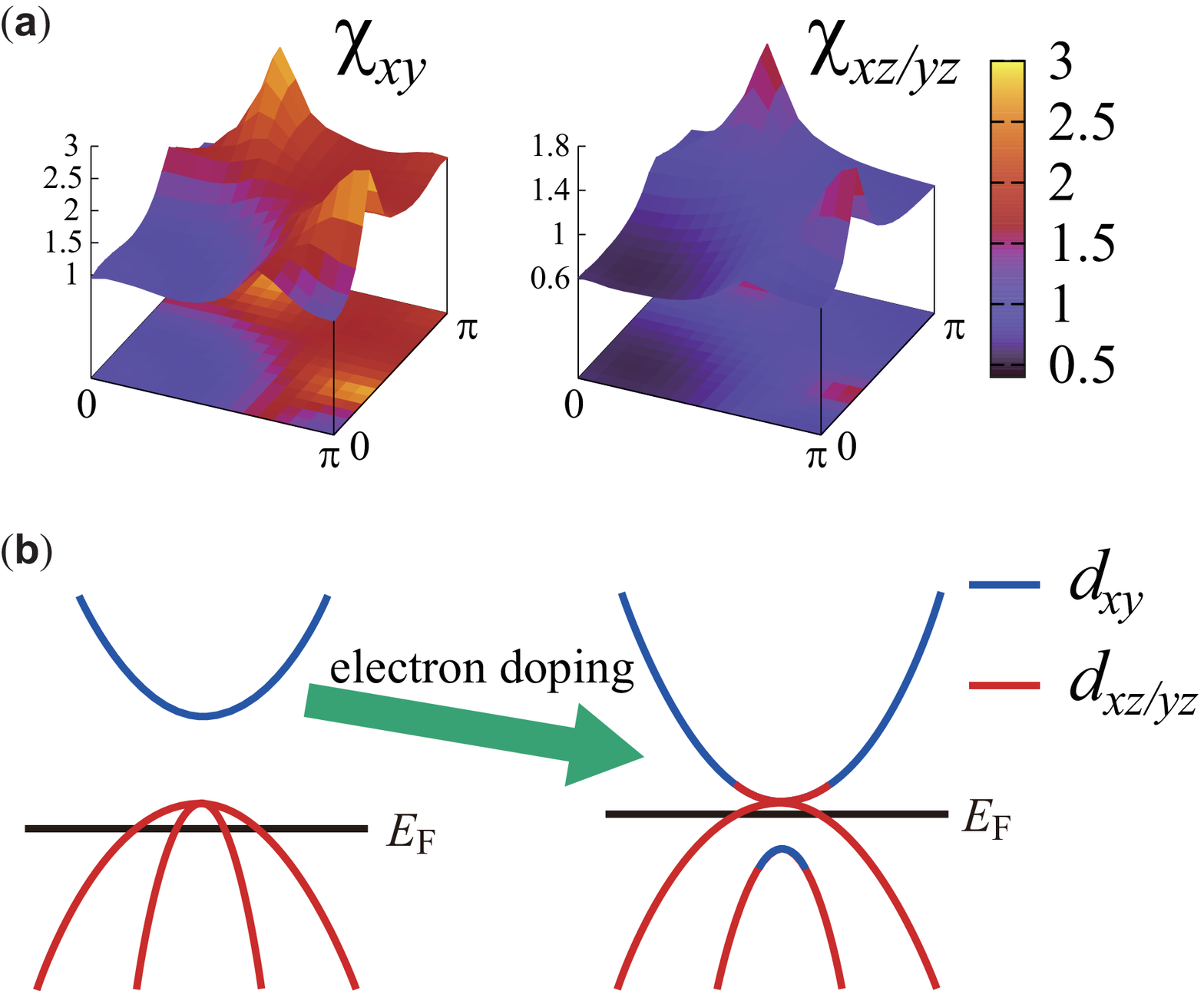}
\caption{(a) Intraorbital spin susceptibilities $\chi_{xy}$ and $\chi_{xz/yz}$ 
for $x=0.35$, $\Delta\alpha=-1^{\circ}$
(b) schematic figure of the band structure variance around (0,0) with doping. 
}
\label{figS5}
\end{figure}

In addition to the re-enhancement of the $d_{xy}$ spin fluctuation, 
the $d_{xz}/d_{yz}$ spin fluctuation is also re-enhanced in a even 
more heavily doped regime as shown in Fig.~\ref{figS5}(a).
This can again be traced back to the band structure variance with doping. 
The $d_{xy}$ portion of the band around $(0,0)$ 
above the Fermi level comes down with doping, and eventually sinks 
below the $d_{xz}/d_{yz}$ bands, as shown in Fig.~\ref{figS2}~(a)$\sim$ (d). 
This is once again due to the reduction of $t_1$.
A band structure reconstruction occurs, and one of the $d_{xz}/d_{yz}$  
bands changes into a concave dispersion as shown schematically in 
Fig.~\ref{figS5}~(b)\cite{Andersen,Miyake42622,UsuiSUST}. 
Therefore, the density of states just above the Fermi level 
(the hole density of states) increases, thereby enhancing the 
electron-hole interaction between $\sim (0,0)$ and $\sim (\pi,0)$.
Reflecting this re-enhancement of the $d_{xz}/d_{yz}$ spin fluctuations, 
there appears a ``kink''  in the $\lambda$ vs. $x$ plot
in the heavily doped regime of $x=0.35\sim 0.45$ (Fig.2 of the main text). 
It should be noted, however, that this effect may be somewhat obscured in the 
actual materials due to the 
three dimensionality of the system, which is neglected here. 
Namely, this band structure reconstruction actually occurs at 
$k_z$ that varies with the doping rate\cite{Andersen,Miyake42622,UsuiSUST}.

\section{Correspondence between real and momentum spaces}

\begin{figure}
\centering
\vspace{1.3em}
\includegraphics[width=9cm,clip]{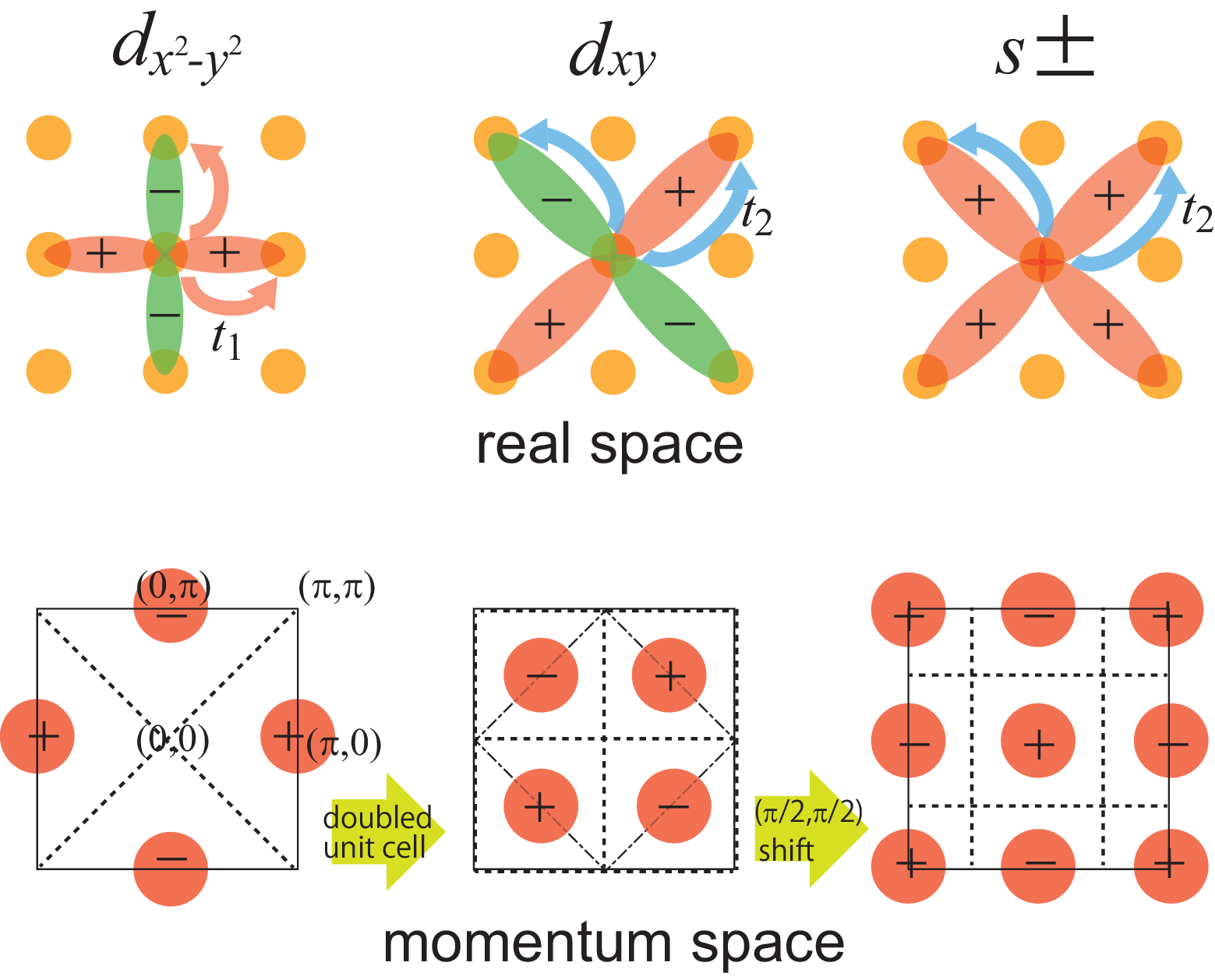}
\caption{Correspondence of pairing in real (upper) and momentum (lower panels) 
spaces. The signs in the upper panels represent the sign of the pair 
wave function. The circled areas in the lower panels are those where the 
density of states near the Fermi energy is large.}
\label{figS6}
\end{figure}

In the cuprates, it is known that the nearest neighbor hopping 
$t_1$ dominates. This is likely to favor nearest neighbor pairing in 
real space, but there are two forms of  nearest neighbor singlet pairing, 
the $d_{x^2-y^2}$ and the 
extended $s$-wave pairings. The former and the latter 
have sign reversing and conserving pair wave functions in real space, 
which corresponds to 
the gap forms $\cos(k_x)-\cos(k_y)$ and  $\cos(k_x)+\cos(k_y)$, respectively, 
in momentum space. In the case of the cuprates, the density of states 
is large around the wave vector $(\pi,0)$ and $(0,\pi)$, and this 
is in favor of the $d_{x^2-y^2}$-wave pairing, whose gap is large around 
those wave vectors. On the other hand, systems with $t_2\gg t_1$ can 
be considered as similar to the cuprates, but with a doubled unit cell, 
if we neglect $t_1$. In momentum space, this gives the gap of the 
form $\sin(k_x)\sin(k_y)$, namely the $d_{xy}$-wave pairing, 
if the density of states is large around the wave vectors 
$(\pm\pi/2, \pm\pi/2)$, which corresponds to $(\pi,0)/(0,\pi)$ 
in the halved Brillouin zone (dash dotted line). 
In real space, $d_{xy}$ pairing is a next nearest neighbor pairing 
whose wave function changes sign with 90 degrees rotation.
Our present study have revealed that 
high $T_c$ iron-based superconductors tend to have large $t_2$, but 
there are no Fermi surfaces around $(\pm\pi/2, \pm\pi/2)$,
which is unfavorable for $d_{xy}$ pairing.
Instead they have Fermi surfaces (or bands near the Fermi level) 
around $(0,0)$, $(\pi,0)$, $(0,\pi)$, $(\pi,\pi)$, 
which are the points shifted by 
$(\pm\pi/2, \pm\pi/2)$ from the wave vectors $(\pm\pi/2, \pm\pi/2)$. 
The $(\pm\pi/2, \pm\pi/2)$ shift transforms the $d_{xy}$ gap into 
$\sin(k_x-\pi/2)\sin(k_y-\pi/2)
=\cos(k_x)\cos(k_y)$ , i.e., the $s\pm$ form. 
This gap corresponds to next nearest neighbor pairing in 
real space, which goes hand in hand with large $t_2$, 
but in a sign conserving form. 
This relation is schematically depicted in Fig.~\ref{figS6}.   

